\documentclass[bibyear]{aa} 
\usepackage{graphicx}
\usepackage{txfonts}
\usepackage{color}

\begin{document}

\title{The relation between radio and X-ray luminosity of black hole
  binaries: affected by inner cool disks?}


\author{E. Meyer-Hofmeister\inst{1} and F. Meyer\inst{1}}
\offprints{Emmi Meyer-Hofmeister; emm@mpa-garching.mpg.de}
\institute
    {Max-Planck-Institut f\"ur Astrophysik, Karl-
     Schwarzschildstr.~1, D-85740 Garching, Germany
     \mail{emm@mpa-garching.mpg.de}
}
\date{Received: / Accepted:}
\abstract
{Observations of the black hole X-ray binaries GX
  339-4 and V404 Cygni have brought evidence of a strong correlation 
  between radio and X-ray emission during the hard spectral state; 
  however, now more and more sources, the so-called `outliers', are
  found with a radio emission noticeably below the established 
  `standard' relation. Several explanations have already been considered,
  but the existence of dual tracks is not yet fully understood.}
{We suggest that in the hard spectral state re-condensation of gas from
  the corona into a cool, weak inner disk can provide additional soft 
  photons for Comptonization, leading to a higher X-ray
  luminosity in combination with rather unchanged radio emission,
  which presumably traces the mass accretion rate.}   
{As an example, we determined how much additional luminosity due to
  photons from an underlying disk would be needed to explain the
  data from  the representative outlier source H1743-322.}
{From the comparison with calculations of Compton spectra with and without
the
  photons from an underlying disk, we find that the required additional X-ray
  luminosity lies well in the range obtained from theoretical models of the
  accretion flow.}
{The radio/X-ray luminosity relation resulting from Comptonization of 
additional photons from a weak, cool inner disk during the hard
spectral state can explain the observations of the outlier sources,
especially the data for H1743-322, the source with the most detailed 
observations. The existence or non-existence of weak inner disks on
the two tracks might point to a difference in the magnetic fields of
the companion stars. These could affect the effective viscosity and
the thermal conductivity, hence also the re-condensation
process.}


\titlerunning {The relation between radio and X-ray luminosity of black hole
binaries}

\maketitle
%


\section{Introduction}

Stellar-mass black hole binaries are sources where gas is accreted
from a companion star either from a massive star, in high-mass X-ray
binaries, or from a star of a few solar masses in low-mass X-ray
binaries (LMXB). The LMXBs are typically transient systems, which are   often
detected
during an outburst and observed mainly during outbursts, but these
binaries have long time intervals of low luminosity between the bright
states. Matter flows continuously from the 
companion star towards the black hole via an accretion disk. During
phases of low luminosity, the accretion disk can be truncated, and in the
inner
regions surrounding the central star, the disk is replaced by an 
advection-dominated radiatively inefficient flow, ADAF (Narayan \& Yi
1994).

 The generally accepted
mechanism that drives the outburst cycles is the ionization
instability in the disk, which was developed earlier in the model for dwarf
nova outbursts (Meyer \& Meyer-Hofmeister 1981). But in X-ray binaries,
the outburst features are more complex owing to the irradiation of the
disk by the innermost hot regions. The enormous variety in the
outburst behaviour of X-ray binaries was already pointed out by Chen
et al. (1997).  In their recent review McClintock \& Remillard (2006)
focus on the 18 black holes with measured masses. The behaviour and the
properties of binaries with dynamically confirmed black holes have been
reviewed by Remillard \& McClintock (2006).

The possible schemes of accretion geometry lead to quite 
different spectra. If the accretion disk reaches inward
towards the last stable orbit, the spectrum is dominated by a strong
thermal component, the state is known as the high/soft state. The
second main spectral state, the low/hard state, is characterized by a power-law
continuum with a spectral index $\Gamma \leq$ 1.8 (Remillard \&
McClintock 2006). Besides these two main states, an
intermediate state and a state of very high accretion rate are known. 
Done et al. (2007) have investigated how a coherent
picture of the physics of the accretion flow onto black holes
can be developed from the increasing number of observations. Dunn et
al. (2010) report on a consistent and comprehensive spectral analysis
of X-ray binaries. Gilfanov (2010) have reviewed the current status of the
theoretical
understanding of accretion and formation of the X-ray
radiation. Investigations of observed spectra of several selected
black hole binaries have been carried
out to study the emission of LMXBs.

New insight comes with the results of the recent {\it{SWIFT}}
survey of all known stellar-mass black hole binaries (Reynolds \& Miller
2013). In comparison to
{\it{RXTE}}, the low-energy X-ray cutoff allows studying the disks at
lower energies, whereas previously a cool inner disk would have been outside
the {\it{RXTE}} low-energy cutoff at ~3keV. This is important for
analyzing inner regions, and it can affect the relation between radio
and X-ray luminosity investigated here. The ejection of relativistic jets
occurs at the transition from the hard spectral state to the
intermediate state during the rise of the outburst (see the `jet line'
in the LMXBs hardness-intensity diagram of Fender et al. 2004).
The radio emission from a different type of jet, a so-called compact
jet, occurs in the hard spectral state, extending to very low
luminosities (for a recent review of radio emission and jets see 
Gallo 2010). The radio emission in 
the hard state is usually characterized by a flat or slightly inverted
spectrum, interpreted as self-absorbed synchrotron emission from a
steady, collimated compact jet. Observations have provided evidence
that a strong connection exists between radio and X-ray emission
during the hard state: an universal correlation of the form
$L_{\rm{radio}} \propto L_{\rm{X}} ^{0.7}$ was found by Gallo, Fender
\& Pooley (2003), dominated by the observations of GX 339-4 and V404 Cyg.

For GX 339-4 a large sample of quasi-simultaneous radio and X-ray
observations were collected during the many outbursts from 1997 to
2011 (Corbel et al. 2013). V404 Cyg was observed during the decay of the
1989 
outburst (Han \& Hjellming 1992; Kitamoto et al. 1990). A definitive
correlation study is presented by Corbel et al. (2008), resulting in
a slight change of the correlation index. The 
correlation appears to be maintained down to the quiescent
level of at least two sources, V404 Cyg, A 0620-00, and possibly also
GX 339-4 (Corbel et al. 2003; Gallo et al. 2003, Gallo et al. 2006).
Important for understanding the accretion geometry in
supermassive black holes, the correlation found for LMXBs, was extended
to active galactic nuclei (AGN) to provide a universal scaling law, which
is the
definition of the `fundamental plane' of black hole activity 
(Merloni et al. 2003; Falcke et al. 2004). It is therefore interesting, also
in this
connection, that recent observations of a number of sources now seem
to indicate that the established correlation is not universal.

During recent years, quite an effort has been made to get
quasi-simultaneous X-ray and radio observations, leading to the result that
for several sources, such as XTE J1550-564, XTE 1650-500, GRO J1655-40,
Swift J1753.5-0127, and H1743-322, for a given X-ray luminosity, 
the simultaneous radio luminosity is lower than that of the standard 
correlation. These sources were classified as
outliers. (Further information can be found for the sources XTE
J1550-564 and GRO J1655-40 in the work by Calvelo et al. (2010), for 
XTE J1650-500 in work by Corbel et al. (2004), for Swift J1753.5-012
and H1743-322 in the papers by Soleri et al. (2010) and Coriat et
al. (2011).)
  In the following we refer to the standard track and the
outliers as upper and lower track, respectively. Gallo, Miller \&
Fender (2012),  
presented a data set of more than hundred simultaneous X-ray and radio 
observations of 18 different black hole X-ray binaries in hard state. 

Since this largest data set clearly documents
the many sources at the 
lower track, we show their data set in our 
Fig.~\ref{f:gallo}. We note a remarkable feature: At some lower value
of radio luminosity, the lower track seems to end, and instead a
transition towards the upper track is indicated. At the upper end
of the considered range, the X-ray luminosity is a few 
percent of the Eddington luminosity. Above this
luminosity the transition to the high/soft state occurs, and the 
compact jet is quenched. Coriat et al. (2011) argue that the steep
high-luminosity part of the lower track can be explained by a
radiatively efficient flow. Employing clustering techniques, Gallo et
al. (2012) came to the conclusion that the observational data of the 
two tracks, the upper and the lower ones, are clearly distinct and 
that there is `no compelling explanation for this behaviour'.

We here present a new suggestion. While mostly one searches for an 
explanation for why for the same X-ray luminosity the radio luminosity is
lower on the lower track, we turn the question around and ask why for
the same radio luminosity the X-ray luminosity is higher on the lower track
than on the upper track. We are motivated here by the perception that the jet
energy and the corresponding radio luminosity originate in the
dominant hard state ADAF-corona, from which the X-ray luminosity takes
away only a smaller fraction. We propose that the increased X-ray
luminosity can be due to additional soft photons from a weak inner
disk. Such an inner disk is sustained by re-condensation of matter from the
corona into the disk, for accretion rates
greater than about $10^{-4} \dot M_{\rm{Edd}}$ (Meyer et al. 2007, Liu
et al. 2007). That the re-condensation does not work at low accretion
rates is important for our suggestion since this could naturally
explain that the lower track ends at a certain luminosity that is related
to
the quiescent state. 

In Sect. 2 we summarize the simultaneous radio and X-ray
observations during hard state, as well as the previously made suggestions
to explain this feature. In Sects. 3 and 4 we refer to earlier work about
the 
existence of an inner disk during the hard state, and we 
discuss how much additional luminosity could arise from this thermal 
component to affect the radio/X-ray correlation. We
compare our results for a higher X-ray luminosity due to additional
photons from an inner weak disk with the observations for H1743-322 in Sect.
5. In Sect. 6 we especially discuss what causes the difference to
the sources on the upper track, which could imply that in the systems
at the standard track with the correlation 
$L_{\rm{radio}} \propto L_{\rm{X}} ^{0.7}$ (or with an exponent 0.6),  no
inner disks are
present.

\section{Observations and previous suggestions}
\begin{figure*}

   \sidecaption
   \includegraphics[width=12cm]{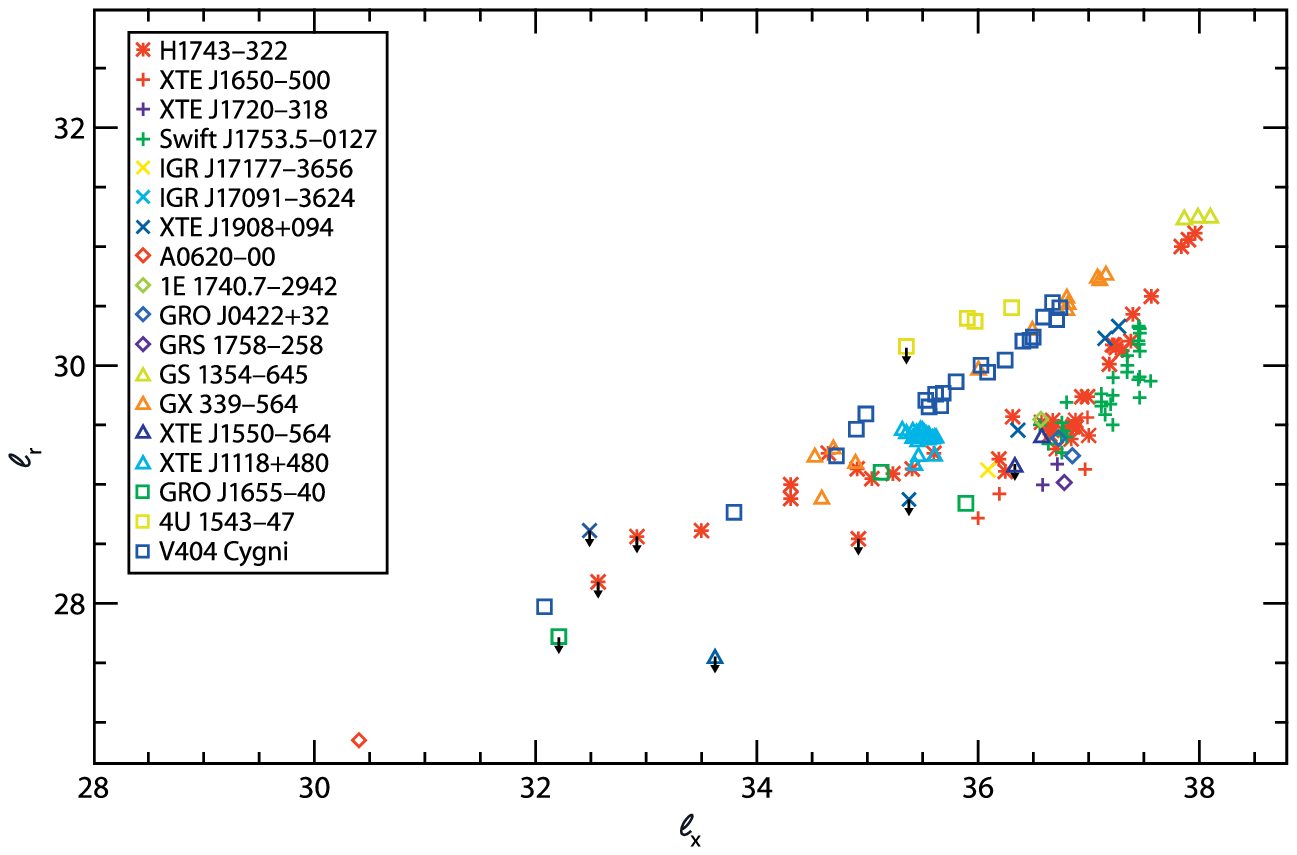}
       \caption{Observed radio and X-ray luminosities of 18 X-ray
         binaries, adopted from Gallo, Miller, and Fender
         (2012, Fig. 1; the original drawing also shows data for Cyg X-1,
         GRS 1915+105 and radio upper limits of quiescent BHXBs from 
         Miller-Jones et al. (2011). Distances of all sources are
         listed and uncertain distances are indicated.)}
         \label{f:gallo}
   \end{figure*}

As mentioned above, the most complete compilation of simultaneous
radio and X-ray observations is by Gallo et
al. (2012). The full data set includes 165 hard-state observations of
18 different X-ray binaries (Fig.\ref{f:gallo}). Some distance 
measurements are uncertain (pointed
out in the original figure), but this cannot explain the two-track
feature. Remarkable is that by excluding the few observations during
quiescence within the range of X-ray luminosities where the two
tracks exist, the values of a chosen source definitely lie on either one
or the other side of the tracks. Ratti et al. (2012) found the radio
luminosity for an additional source XTE J1752-223, on the lower
track. A second recent source, MAXI J1659-152 (Jonker et
al. 2012, van der Horst 2013), also was found close to the lower track.

{Already in 2003 the tight
correlation between radio and X-ray flux was found for GX 339-4 (Corbel
et al. 2003) and for  V404 Cyg (Gallo et
al. 2003). While for the latter source all information comes from one
outburst (decline) in 1989, for GX 339-4 a large sample of data
includes observations over a 15- year period (Corbel et al. 2013). 
All these radio and 
X-ray luminosities lie on the standard correlation. The lower track is
populated by data from a large number of X-ray binaries (see 
Fig.\ref{f:gallo}). The source XTE J1118+480 is sometimes listed under
the outliers, but the luminosity values are relatively high.

Coriat et al. 2011 present the most detailed information with simultaneous radio and X-ray
observations for seven outbursts of H1743-322 between 2003 and
2010. In their paper the observed radio and X-ray
flux densities are listed for all
observations and the position of the particular observation in the
respective hardness-intensity
diagram of that outburst is shown. We discuss these values in our comparison
of theoretical results and observations in Sect. 4.
 -  
Several recent investigations have focused on the question of what might
cause the two different relations between radio and X-ray luminosity. Gallo et
al. (2012) summarize these suggestions. The effects
discussed are hysteresis behaviour (Russell et al. 2007), a large scatter due
to differences in the strength of the magnetic field from source to
source (Casella \& Pe'er 2009, Pe'er \& Casella 2009), different slopes
of the tracks, which otherwise are not expected within the jet-accretion models
(Markoff et al. 2001, 2003), or the low and high spin of the
black hole (Fender et al. 2010). Coriat et al. (2011) suggest 
that the steep part of
the lower track can be understood in the framework of radiatively
efficient accretion,  and the outflow properties of the jet were 
also considered as a possible cause for the two tracks. Soleri 
\& Fender (2011) discuss whether the parameters of the binary 
systems (orbital period, disk size, inclination), as well as the
outburst properties (e.g. low/hard state only outbursts), show any
correlation with the energy output of the jet, but did not find an
obvious dependence.

\section{The existence of cool inner disks during the hard state}
The configuration of the accretion flow towards a black hole,
which depends on the accretion rate, as displayed by Esin
et al. (1997), is commonly accepted. (The observed hysteresis
in the spectral state transition can be understood as an irradiation
effect of the hard state accretion, Meyer-Hofmeister et al. 2005.)   
Still under debate is the disk
truncation. A new discussion arose a few years ago, initiated by new 
observations that indicate a weak soft thermal component in the
spectra of GX 339-4 (Miller et al. 2006a) and SWIFT J1753.5-127 
(Miller et al.2006b), for example. Weak thermal components had already been marginally found
for some sources earlier, always at a time close to an
intermediate spectral state. It is clear that in outburst
decline, at the time when the spectral transition happens, the accretion 
geometry might be complex. The change from disk 
accretion to accretion via a hot flow occurs when the mass flow in the
disk becomes low, lower than the evaporation rate, so that all matter
in the disk is evaporated and flows farther inward as
coronal hot gas (Meyer et al. 2000a, 2000b). This might first happen
at that distance from the black hole where evaporation is most
efficient, e.g. at a hundred or more Schwarzschild radii, depending on
such parameters as viscosity and the ratio of magnetic pressure to total
pressure (Qian et al. 2007). With more
decreasing mass flow, the percentage of coronal flow compared to
the total flow increases, and less gas is accreted via the inner part of the
still co-existing disk. The remaining inner disk would rapidly 
disappear, if not sustained by re-condensation of gas from the corona
into the disk (Meyer et al. 2007, Liu et al. 2007). The
re-condensation of gas from the ADAF into the inner disk follows from
thermal conduction between disk and hot flow. 

Such an innermost disk is
an essential feature of the accretion geometry during
the change from the soft to the hard spectral state. The innermost
disk is weak and thus clearly distinguishable from a standard disk in soft
state. Its appearance is not in conflict with the basic picture of 
disk truncation as sometimes surmised.

An underlying disk provides soft photons for the Comptonization by
electrons in the hot advection-dominated flow. The emission from the
inner cool disk is partly at very low energies, below 0.1 keV, therefore 
not detectable with ${\it{RXTE}}$ with the low-energy
cutoff at $\sim$ 3~keV. In the past, observational data were classified as
during spectral hard state if no accretion disk
component was needed to correctly fit the data above 3~keV,
besides a power-law photon index $\Gamma \leq 2$ (Coriat et al. 2011).
This means that these `hard state' data sources used for the
radio/X-ray correlation might well have additional low-energy thermal 
radiation. 

In their very recent review, Reynolds \& Miller (2013) present a 
{\it{SWIFT}} survey of all stellar mass black hole binaries which
provides a
sample of 476 X-ray spectra, which allows studying the disk
evolution over a very wide range in flux and temperature. The 
sensitivity allows observations of accretion disks down to 
$10^{-3} L_{\rm{Edd}}$. It is noteworthy that for the sample with an
average luminosity of about 1\% Eddington, an accretion disk was
detected in 61$\%$ of the observations. (For the fit of the hard
component a simple power-law model and a Comptonization model were considered.)
This sample is the largest data set of such cool accretion disks studied to
date.

\section{Additional luminosity from a cool inner disk - 
          the effect on the radio/X-ray luminosity correlation}
Scattering of soft photons from an underlying disk adds to the scattering
of synchrotron and bremsstrahlung photons from the ADAF. In their
analysis of the relation between the photon index and the
Eddington ratio, Qiao \& Liu (2013) calculated 
spectra that includes the radiation of an innermost accretion disk. The
strength of the inner disk is determined according to possible
re-condensation of gas from the corona to the disk below. The
re-condensation process works as long as the mass flow
rate in the corona is high enough, $\ge 10^{-4}$ to $10^{-3} L_{\rm{Edd}}$.
The detailed results depend on parameters such as viscosity and
the ratio of magnetic pressure to gas pressure. But despite the 
uncertainties, the calculated spectra show that
for mass accretion rates around 2\% of the Eddington rate, the
underlying disk contributes almost twice as much to the radiation as
the coronal flow alone in the 3-9 keV range, and even more in the 
0.2-2 keV range (Qiao \& Liu, Fig. 3). 

To study the effect on the radio/X-ray luminosity relation, we
take the values 0.0025, 0.005, and 0.01 
$\cdot L_{\rm{Edd}}$ in the following for an
assumed additional luminosity caused by the photons of an underlying
disk. We compare the resulting increase in luminosity with the
observations for H1743-322.

\section{Comparison with observations for H1743-322}

 \begin{figure*}
 \sidecaption
 \includegraphics[width=12.0cm]{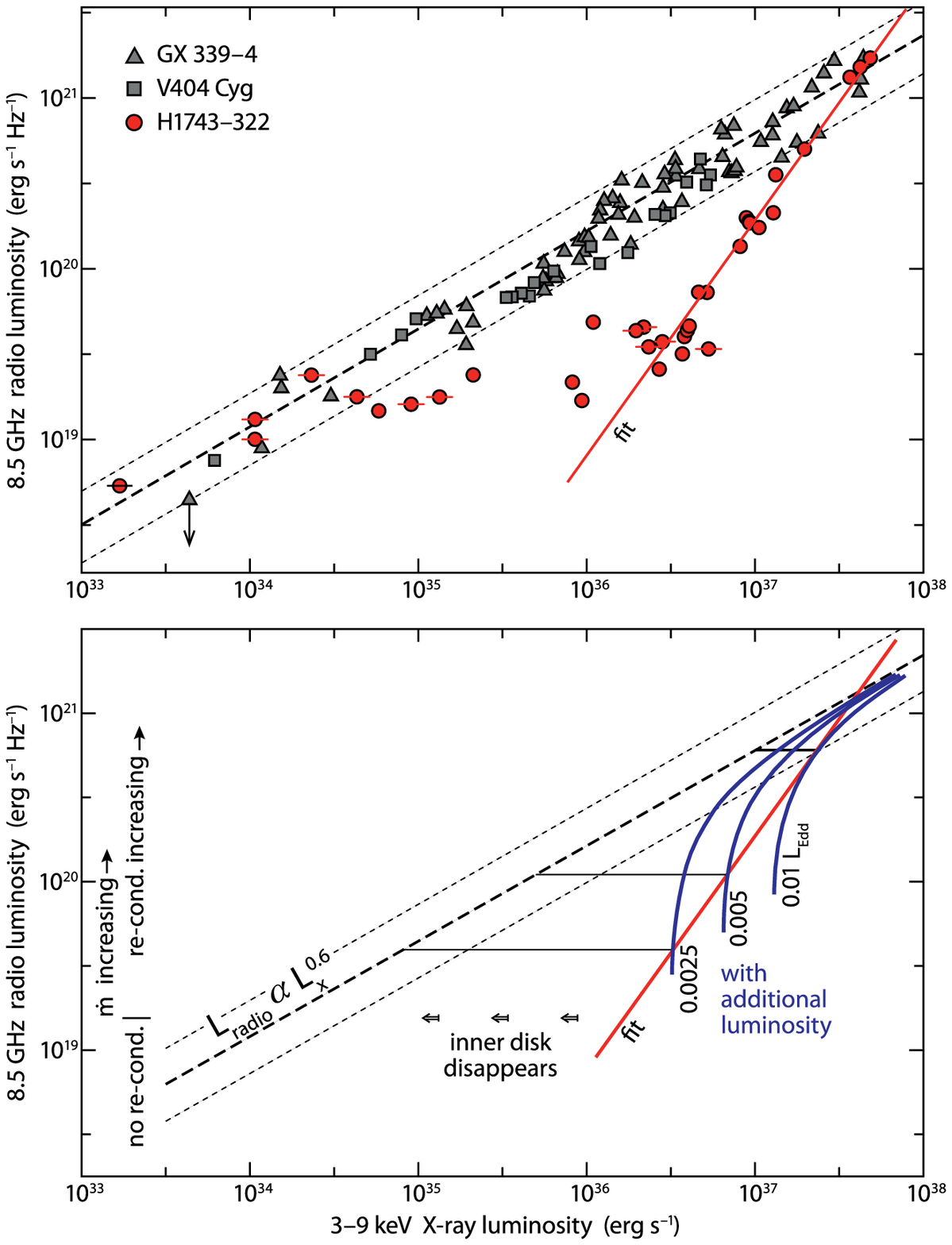}
       \caption{Upper panel: figure adopted
         from Coriat et al. (2011, Fig.5): `standard' correlation
         between  radio and X-ray luminosity $L_{\rm{radio}} \propto L_{\rm{X}} ^{0.6}$,
         defined by the black hole binaries GX 339-4 and V404 Cyg
         (dashed lines indicate dispersion)
         and data for H1743-322, typical of
         the so-called outliers on the lower
         track. Red line: fit to the high luminosity data of
         H1743-322; data for the outburst 2008a 
         marked by horizontal slashes.\,\,\,\,
          Lower panel:  `standard' correlation between radio and 
         X-ray luminosity and fit to the data for H1743-322 as in upper
         panel; blue lines: X-ray luminosity including additional
         luminosity caused by the photons of an
         underlying disk, $\Delta L_{\rm{X}}/L_{\rm{Edd}}$=
         0.0025, 0.005 and 0.01); horizontal lines: the increase
         required for the red line fit to the data of H1743-322.}
        \label{f:lrlx}
   \end{figure*}

 The upper panel of Fig.~\ref{f:lrlx} shows the observations for H1743-322,
together with observations for GX 339-4 and V404 Cyg in the hard
state, adopted from the long-term study of Coriat et al. (2011,
Fig.5). The observations for H1743-322 include
seven outbursts between 2003 and 2010. Only for the outburst in 2003 do
the observations cover a full outburst cycle with rise and
decline including the state transitions from hard to soft and back. For the
remaining outbursts, with several only in hard state, mainly data during 
decline are available. Coriat et al. (2011) fitted the
high-luminosity data of H1743-322, and in the diagram the fit is shown as a red
line. The data were interpreted by the authors as resulting from a radiatively 
efficient hot flow, LHAF (luminous hot accretion flow, see Yuan
2001).                        

In the lower panel of Fig.~\ref{f:lrlx}, the effect of the contribution
from an underlying cool disk is shown. We take the radio luminosity as
a tracer of the mass accretion rate $\dot M$. For any given mass
accretion rate, i.e. for radio luminosity, we consider the increase in the
X-ray luminosity $\Delta L_{\rm{X}}/L_{\rm{Edd}}$ caused by the photons from the
 underlying disk. In the figure we show a grid of positions for
 different additional luminosities (blue lines). We use this grid
 to determine how much $\Delta
 L_{\rm{X}}/L_{\rm{Edd}}$ is required to find the system at the position of the
 observed radio/X-ray correlation of H1743-322, the various positions
 at different radio luminosities given by the red line. (The thin black lines
 connect the $L_X$ values without and with additional
 luminosity.) 

The shape of the red track shows that the required additional
 contribution of the inner disk increases with the  X-ray
 luminosity, which corresponds to the mass flow rate in the
 corona, as predicted by the re-condensation model (Meyer et al. 2007).
 From the comparison of the observations represented by the fit line with
 the grid of blue lines we
 find that the values $\Delta L_{\rm{X}}/L_{\rm{Edd}}$ needed to
 obtain the position at the outlier track lie in the range 0.0025 to
 0.01 $L_{\rm{Edd}}$. The lower the original radiation (from scattering the 
synchrotron and bremsstrahlung photons in the corona), the more
prominent the effect of photons from the underlying disk. Such
luminosity values are plausible within the context of the
re-condensation 
model, which describes the accretion geometry during that spectral state
(Sect. 4).

How can the observations at lower luminosity be understood? For the
X-ray luminosities in the range $10^{35}$ to $10^{36} \rm{erg s^{-1}}$, the radio
emission does not change much. Coriat et al. (2011) interpret this
range as a range of transition from an LHAF to an ADAF
(indicated in their Fig.7). In our model, re-condensation 
is only expected for mass accretion rates that are not too low, above
$10^{-4}$ to $10^{-3} L_{\rm{Edd}}$, that is for
radio luminosities above a few times $10^{19} \rm{erg s^{-1}
  Hz^{-1}}$ (the range indicated on
the left side of Fig.~\ref{f:lrlx}, lower panel).
The X-ray luminosity decreases when the inner disk is not sustained by 
re-condensation. To show the evolution of the luminosity during an
outburst, we mark the data of the outburst 2008a 
with an additional slash. While the 8.5 GHz radio flux density
decreased from 0.23 to 0.13 mJy, the X-ray flux changed from 1.74 to 0.14 
$10^{-11} \rm{erg s^{-1} cm^{-2}}$  (Coriat et al. 2011, Table 1). The
lowering to about 1/10 in X-rays within 12 days can be understood as
due to the disappearance of the innermost disk.
 
Since for very low accretion rates, an innermost disk can no longer be
sustained by re-condensation, then only one track of
radio/X-ray luminosity correlation should be found, a continuation of the
upper track down to quiescence, in agreement with observations. 
At high luminosities, close to those of the transition to the soft
state, the mass accretion rate is so high that the effect of an
underlying disk becomes unimportant, and also the lower track joins the
upper track there.

\section{Discussion}

\subsection{Inner disks and X-ray spectral-timing analysis}

Since the inner accretion disk in the hard state is an essential
feature of our model for explaining the radio/X-ray
luminosity relation, the question arises as to how such an inner disk would
be recognized in the spectral-timing analysis.
This analysis provides another view of the accretion geometry besides
the widely discussed picture of a truncated outer disk,
together with an ADAF in the inner region, understood theoretically
(Narayan et al. 1998) and confirmed by observations for many LMXBs. 
Bringing these aspects together might lead to the following picture of
the accretion flow.

The observed variability of the power-law continuum in hard state shows
spectral-timing properties that can be attributed to accretion rate
fluctuations (Kotov, Churazov \& Gilfanov 2001, Ar\'evalo \& Uttley 2006).
A new spectral analysis technique, the `covariance spectrum method'
allows studying how the contributions of the components of the  
spectrum and variations on different timescales are related
(Wilkinson \& Uttley 2009; Uttley et
al. 2011). Observed spectral time lags of GX 339-4 and other sources
make it clear that variations in the disk black body emission
substantially lead variations in the power-law emission, which is
consistent with the geometry of a soft component arising farther outwards than
the hard component.} From their analysis of observations of GX 339-4 
and SWIFT J1753.5-0127 on 2004 March 17 and 24,  Wilkinson \& 
Uttley concluded that at these times the disk truncation would have
been less than 20 Schwarzschild radii. For both observations, the 
{\it{XMM-Newton}} spectra reveal the presence of a cool
accretion disk (Miller et al. 2006a, 2006b). In the
re-condensation interpretation, this would be in agreement with the
weak inner disk and would not impinge on a possible truncation of the
standard disk at large radii. The observation for
GX 339-4 seems to be in an almost intermediate spectral state as discussed in Sect. 6.2.

If we consider these variations in the outer disk and the consequential
variability in the inner hot flow, we expect varying re-condensation
that feeds an inner disk. The varying mass flow in the inner disk
propagates by diffusing inwards, leading to the observed time lags between
outer (cooler) and inner (hotter) regions of this inner disk. The
innermost regions of this inner disk provide the most effective
photons for Comptonization in the ADAF/corona. This contribution to the
power law besides the synchrotron and bremsstrahlung photons will then
vary in phase with the innermost disk photons and show the observed
lags to the cooler photons from the outer part of the accretion disk.

\subsection{The standard correlation}

The upper track of the radio/X-ray correlation is dominated by 
GX 339-4 and V404 Cyg. The source GX 339-4 was observed many times in
hard state. During the long-term campaign of Corbel et al. (2013),
quasi-simultaneous radio and X-ray observations in hard state were
collected, all corresponding to the standard track. \it{
{SWIFT}} \rm{and} {\it{RXTE}} \rm{observations of GX 339-4 in the hard
  state by {Allure et al.} (2013) provide information on the
different truncations of the accretion disk during three time intervals between
2006 and 2010. Plant et al. (2013) in their recent
work found disk truncation, also in GX 339-4, at distances of hundreds
of Schwarzschild radii for very low-luminosity/low- accretion rate and
near the ISCO for higher accretion rate, the latter representing a
hard intermediate state.

According to our suggestion no inner disks should exist during the
hard state of sources on the upper track. There is} \rm{ one special
observation for GX 339-4: Miller et
al. (2006b) find thermal emission from an accretion disk in the
low/hard state in the rising phase of the 2004 outburst,
interpreted as a weak inner disk. But the flux
observed on 2004 March 16/17 is close to the highest values found
during the rise. The increasing luminosity is already about 0.05
$L_{\rm{Edd}}$ (Corbel et al. 2013, Table 1), so that the state seems to
be an advanced intermediate spectral state, close to the spectral 
transition and in the radio/X-ray diagram close to the upper merging
of the two tracks. In that situation the contributions of the observed
soft photons to the Comptonization of the corona is probably small
compared to that of synchrotron and bremsstrahlung photons of the ADAF.

\subsection{A systematic difference between the upper and lower track?}
The difference between the two tracks seems to require a systemic
difference between sources of otherwise similar black hole mass and
accretion rates. There appears to be little alternative to locating this
difference in the secondary stars and then in their magnetic fields. These
fields can reach over and be embedded in the accretion disk and even
be concentrated by the accretion flow. Such fields could indeed influence
the re-condensation process and thereby the existence and strength
of an inner disk. For example, good thermal conduction requires a rather
direct magnetic field connection between the disk and the lower
regions of the corona above which the material
recondenses. Secondly analysis of the re-condensation physics yields
a strong dependence on the frictional parameter $\alpha$ in the
corona, which will also depend on the magnetic field.

With such a picture in mind, the companion stars of the sources on the
two tracks are of interest. Stars with lower mass 
than the sun have a convective zone in the outer part and are
radiative in the inner part, operating a dynamo at the base
of the convective zone. More massive stars are radiative in the outer
regions, convective inside. The A stars in this group show no strong magnetic
fields, but the Ap stars, a subgroup, show strong magnetic fields and
possibly have magnetic fields in channels through the adiabatic region,
confined in the convective core (Meyer 1993). One expects the same
phenomenon in all the more massive stars.

The orbital periods of the companion stars give a rough estimate of
the mass of the Roche-lobe filling secondary star. Soleri \&
Fender (2011) have tested in detail whether, for a sample of 17 black
hole binaries, a connection exists between the energy output of the jet and
the orbital period and other characteristic parameters such as the size
of the disk, the inclination, and the outburst properties. They did
not find any association. However, we note that all short-period systems
(i.e. secondary stars of low mass) in that sample lie on the lower
track (see Corbel et al. 2013, Fig.~9), only the position of XTE
J1118+480 is somewhat unclear. The two recent sources XTE J1752-233
(Ratti et al. 2012) and MAXI J1659-152 (Jonker et al. 2012) both are on
the lower track and have short orbital periods, 6.8\,h and 2.4\,h. The 
sources with long
orbital periods (i.e. more massive secondary stars), 26.8 to
155.3 hours (periods from Ritter \& Kolb 2003), are mainly on the
upper track, but a few are on the lower track, as one would expect from
the magnetic dichotomy in the more massive stars. 
It seems surprising that different magnetic field strengths from the
secondary stars do not lead to significant scatter around the lower
track. This might perhaps indicate that the dominant effect of the
magnetic fields on the re-condensation process is the establishment
or non-establishment of a good magnetic connection between lower
corona and disk, as required for thermal conduction. 

The question arises whether the magnetic difference
of the sources would not also be mirrored in a difference between the
radio luminosity of their jets. However, the cut-off of the lower track and the 
transition to the upper track at a nearly constant radio luminosity
seems to indicate that such an effect is rather small.

\subsection{The energy output of the jet affected by an underlying disk?}

During the hard to soft transition a drop in radio
flux is observed for many sources and attributed to the quenching of the
compact (core) jet, such as for GX 339-4 in the high/soft
X-ray state by a factor around 25 in comparison with the low/hard
state (Fender et al. 1999). Quenching factors that are even much larger, up
to several hundred, were found for H1743-322 (Coriat et al. 2011) and 4U 1957+11 (Russell et al. 2011).

The theoretical modelling of jet power can explain the large
difference between hard and soft states. Meier (2001) found a jet power 
from a thin disk to be 100 times weaker than in the ADAF case. The square of
the half thickness of the disk enters into the jet power
formula. Meyer-Hofmeister et al. (2012) investigated the
structure of coronae above accretion disks and the change of ion
temperature with a decrease in electron temperature due to Compton cooling.
For low accretion rates $\dot m$ between $10^{-3}$ to $10^{-2}$, the
ion temperature is affected only after a strong decrease in electron 
temperature. Since the scale height depends on the ion temperature, this 
means little change in the half thickness
of the ADAF/corona in the inner region, from which matter
re-condensates into the weak disk. The models and simulations of
jet production (Blandford \& Znajek 1977; Blandford \& Payne 1982; 
Meier et al. 1997) show that it is the
strength of the poloidal component of the magnetic field that is
largely responsible for producing the jet. Now, if the
accretion geometry with a weak inner disk is only
slightly different from a pure ADAF, one expects only a minor
effect on the jet power. A reduction of the jet power would lead to
lower radio emission for a given X-ray luminosity, which is the same
effect on the correlation as from the photons of an inner disk, but 
probably much weaker.

\section {Conclusions}

After more and more hard X-ray sources were observed
whose radio luminosity was lower than expected from the standard 
correlation (Gallo et al. 2003) and after it became clear that the two 
tracks of the radio/X-ray correlation are distinct (Gallo
et al. 2012), we suggest that the existence or non-existence of a
weak inner disk could provide a physical explanation.

If a weak inner disk exists during the hard
spectral state, the Comptonization of the disk photons in the corona 
provides additional hard X-ray emission. Such a disk is sustained by 
re-condensation of coronal matter (Meyer et al. 2007; Liu et
al. 2007). This shifts the X-ray luminosity to values higher than
those expected from an inner ADAF alone. As the radio luminosity of the 
compact jet emanating from the inner ADAF/corona traces $\dot m$ and
the X-ray luminosity is increased, the radio/X-ray luminosity
correlation becomes steeper, as found for the outlier source H1743-322 displayed
in Fig.\ref{f:lrlx}. The red line shows the fit to these high-luminosity data (Coriat et al. 2011), representative of the lower track.

The comparison between our model and the observations 
yields the following results: \\
(1) To get the correlation between radio and X-ray flux (described by
the red fit line), an additional luminosity of 
about 0.05$L_{\rm{Edd}}$ from Comptonization of photons from a weak
inner disk would be required. This order of magnitude is suggested by
computations of spectra that include an inner weak disk in hard spectral
state (Qiao \& Liu 2013).\\ 
(2) The re-condensation model predicts that no inner disk can exist
for accretion rates below a certain limit somewhere around $10^{-3}$
to $10^{-4}L_{\rm{Edd}}$. This can help for understanding
why the lower track of outliers ends in a transition to the
standard track at low luminosity. \\ 
(3) At high luminosity where the systems approach the soft state the 
additional luminosity becomes unimportant and consequently the two
tracks merge.

In the radio/X-ray luminosity diagram, the data points for neutron stars lie near
the lower track (Soleri \& Fender, Fig. 6). It would be interesting to
clarify whether their position can be understood as due to the
additional luminosity from the neutron star surface that provides
photons for a coronal Comptonization. For the chosen sample of neutron stars in their analysis, Migliari \& Fender (2006) find
 a steeper
slope in the radio/X-ray relation than the one known for black hole
binaries.

  A confirmation of our model could come from analysis of
spectra in the low energy range, such as \it{SWIFT}}
spectra, with simultaneous radio observations. If further support for
the weak inner disks is found, it would
be interesting to consider the situation in AGN and consequences for
the `fundamental plane' for black hole activity (Merloni et al. 2003).

\begin{acknowledgements} 
{We thank Marat Gilfanov for discussions and the referee for many
  helpful comments.} 
\end{acknowledgements}

{}


\begin{thebibliography}{}

\bibitem{} Allured, R., Tomsick, J.A., Kaaret, P. et al. 2013, arXiv:
  1307.3503
\bibitem{} Ar\'evalo, P. \& Uttley, P. 2006, MNRAS 367, 801 
\bibitem{} Blandford, R.D., \& Payne,D.G. 1982, MNRAS 199,883
\bibitem{} Blandford, R.D., \& Znajek,R. 1977, MNRAS 179, 433
\bibitem{} Calvelo, D.E., Fender, R.P., Russell, D.M. et al. 2010,
  MNRAS 409, 839
\bibitem{} Casella, P.G., \& Pe'er, A. 2009, ApJ 703, L63
\bibitem{} Chen, W., Shrader, C.R., \& Livio, M. 1997, ApJ 491, 312
\bibitem{} Corbel, S., Nowak, M.A., Fender, R.P. et al. 2003, A\&A
  400, 1007
\bibitem{} Corbel, S., Fender, R.P., Tomsick, J.A. et al. 2004, ApJ
  617, 1272 
\bibitem{} Corbel, S., Koerding, E. \& Kaaret, P. 2008, MNRAS 389,
  1697 
\bibitem{} Corbel, S., Coriat, M., \& Brocksopp, C. 2013, MNRAS 428, 2500
\bibitem{} Coriat, M., Corbel, S., Prat, L. et al. 2011, MNRAS 414, 677
\bibitem{} Done, C., Gierlinski, M., \& Kubota, A. 2007,
  Astron. Astrophys. Review, Vol. 15, p.1
\bibitem{} Dunn, R.J.H., Fender, R.P., K\"ording et al., 2010, MNRAS 403,
  61
\bibitem{} Falcke, H., Koerding, E., \& Markoff, S. 2004, A\&A 414, 895
\bibitem{} Fender, R.P., Corbel, S., \& Tsoumis, T. 1999, ApJ 519, L165
\bibitem{} Fender, R.P., Belloni, T.M., \& Gallo, E. 2004, MNRAS 355,
  1105
\bibitem{} Fender, R.P., Gallo, E., \& Russell, D.M. 2010, MNRAS 406,
  1425
\bibitem{} Gallo, E. 2010, in The Jet Paradigm - From Microquasars
  to Quasars, Lecture Notes in Physics 794 (Springer), 85
\bibitem{} Gallo, E., Fender, R.P., \& Pooley, G.G. 2003, MNRAS, 344, 60 
\bibitem{} Gallo, E., Fender, R.P., Miller-Jones, J.C.A. et al. 2006,  
  MNRAS 370, 1351)
\bibitem{} Gallo, E., Miller, B.P., \& Fender, R. 2012, MNRAS 423, 590
\bibitem{} Gilfanov M. 2010, in The Jet Paradigm - From Microquasars
  to Quasars, Lecture Notes in Physics 794 (Springer), 17
\bibitem{} Han, X., \& Hjellming, R.M. 1992, ApJ 400,
304
\bibitem{} Jonker, P.G., Miller-Jones, J., Homan, J. et al. 2010,
  MNRAS 401, 1255
\bibitem{} Jonker, P.G., Miller-Jones, J., Homan, J. et al. 2012,
    MNRAS 423, 3308 
\bibitem{} Kitamoto, S., Tsunemi, H., Pedersen, H. et al. 1990, ApJ 361,
590
\bibitem{} Kotov, O., Churazov, E., \& Gilfanov, M. 2001, MNRAS 327,
  799
\bibitem{} Liu, B. F., Meyer, F., \& Meyer-Hofmeister, E. 2006, A\&A  
  454, L9                                                             
\bibitem{} Liu, B. F., Taam, R.E., Meyer-Hofmeister, E. et al. 2007,
  ApJ 671, 695  
\bibitem{} Markoff, S., Falcke, H., \& Fender, R. 2001, A\&A 372, L25
\bibitem{} Markoff, S., Nowak, M., Corbel, S. et al. 2003, A\&A 397, 645
\bibitem{} McClintock, \& J.E., Remillard, R.A. 2006, in Compact 
Stellar X-ray Sources, ed. W.H.G. Lewin, M. van der Klis,
Cambridge Astrophysics Series No. 39 (Cambridge University Press), 157      
\bibitem{} Meier, D.L. 2001, ApJ 548, L9   
\bibitem{} Meier, D.L., Edgington, S., Godon, P. et al. 1997, Nature
  388, 350
\bibitem{} Merloni, A., Heinz, S., \& di Matteo, T. 2003, MNRAS 345, 1057
\bibitem{} Meyer, F. 1993, in Cosmical Magnetism, NATO ASI Series,
  Vol. 422 (Kluwer Academic Publishers), 67
\bibitem{} Meyer, F., \& Meyer-Hofmeister, E. 1981, A\&A 104, L10
\bibitem{} Meyer, F., Liu, B.F., \& Meyer-Hofmeister, E. 2000a, A\&A 354,
\bibitem{} Meyer, F., Liu, B.F., \& Meyer-Hofmeister, E. 2000b, A\&A 361,
  175
\bibitem{} Meyer, F., Liu, B.F., \& Meyer-Hofmeister, E. 2007, A\&A 
  463, 1
\bibitem{} Meyer-Hofmeister, E., Liu, B.F., \& Meyer, F. 2005, A\&A
  432, 181                  
\bibitem{} Meyer-Hofmeister, E., Liu, B.F., \& Meyer, F. 2012, A\&A 544,
  87
\bibitem{} Migliari, S. \& Fender, R.P. 2006, MNRAS 366, 79
\bibitem{} Miller, J.M., Homan, J., \& Miniutti, G. 2006a, ApJ 652, L113
\bibitem{} Miller, J.M., Homan, J., Steeghs et al. 2006b, ApJ 653, 525  
\bibitem{} Narayan, R., \& Yi, I. 1994,  ApJ 428, L13  
\bibitem{} Pe'er, A., \& Casella, P.G. 2009, ApJ 699, 1919
\bibitem{} Plant, D.S., Fender, R.P., Ponti, G. et
   al. arXiv:1309.4781
\bibitem{} Qian, L., Liu, B.F., \& Wu, X.-B. 2007, ApJ 668, 1145
\bibitem{} Qiao, E., \& Liu, B.F. 2013, ApJ 764, 2
\bibitem{} Ratti, E.M., Jonker, P.G., Miller-Jones, J.C.A. et
  al. 2012, MNRAS 423, 2656
\bibitem{} Remillard, R.A., \& McClintock, J.E. 2006, ARA \& A 44, 49   
\bibitem{} Reynolds, M.T., \& Miller, J.M. 2013, ApJ 769, 16
\bibitem{} Ritter, H., \& Kolb, U. 2003, A\&A 404, 301 
\bibitem{} Russell, D.M., Maccarone, T.J.,K\"ording, E.G. et al. 2007,
\bibitem{} Soleri, P., \& Fender, R. 2011, MNRAS 413, 2268  MNRAS 379,
  1401
\bibitem{}  Soleri, P., Fender, R., Tudose et al. 2010, MNRAS 406,
   1471
\bibitem{} Taam, R.E., Liu, B.F., Meyer, F. et al. 2008, ApJ 688, 527
\bibitem{} Uttley, P., Wilkinson, T., Cassatella, P. et al. 2011, MNRAS
  414, L60
\bibitem{} van der Horst, A.J., Curran, P.A., Miller-Jones, J.C.A. et
  al. 2013, MNRAS 436, 2625 
\bibitem{} Wilkinson, T., \& Uttley, P. 2009, MNRAS 397, 666 
\bibitem{} Yuan, F. 2001, MNRAS, 324, 119


\end{thebibliography}
\end{document}